\documentclass[10pt,journal,compsoc]{IEEEtran}
\IEEEoverridecommandlockouts
\usepackage{cite}
\usepackage{amsmath,amssymb,amsfonts}
\usepackage{algorithmic}
\usepackage{graphicx}
\usepackage{textcomp}
\usepackage{xcolor}
\usepackage[utf8]{inputenc}
\usepackage{comment}
\usepackage[thinlines]{easytable}

\usepackage{slashbox}
\usepackage{balance}
\usepackage{times}

\usepackage[labelformat=simple, labelsep=colon]{subcaption}
\DeclareCaptionSubType*{figure}

\def\BibTeX{{\rm B\kern-.05em{\sc i\kern-.025em b}\kern-.08em
    T\kern-.1667em\lower.7ex\hbox{E}\kern-.125emX}}
\begin{document}

\title{Cyberphysical Blockchain-Enabled Peer-to-Peer Energy Trading}

\author{
\IEEEauthorblockN{\textbf{Faizan Safdar Ali}, \textit{Student Member, IEEE}, \textbf{Moayad Aloqaily},
\textit{Member, IEEE}, \\\textbf{Omar Alfandi}, \textit{Member, IEEE}, and \textbf{\"Oznur \"Ozkasap}, \textit{Senior Member, IEEE}
} 
\IEEEcompsocitemizethanks{

\IEEEcompsocthanksitem Faizan Safdar Ali and \"Oznur \"Ozkasap are with Department of Computer Engineering, Ko\c{c} University, Istanbul, Turkey. \protect E-mails: \{fali18, oozkasap\}@ku.edu.tr

\IEEEcompsocthanksitem M. Aloqaily is with xAnalytics Inc., Ottawa, ON, Canada. \protect E-mail: maloqaily@ieee.org

\IEEEcompsocthanksitem Omar Alfandi is with Georg-August-University Goettingen, Germany. \protect E-mail: alfandi@cs.uni-goettingen.de

}
}


\maketitle
\begin{abstract}

Scalability and security problems with centralized architecture models in cyberphysical systems have provided opportunities for blockchain-based distributed models.
A decentralized energy trading system takes advantage of various sources, and effectively coordinates the energy to ensure optimal utilization of available resources. It achieves this by managing physical, social and business infrastructures, using technologies such as the Internet of Things (IoT), cloud computing and network systems. 
To address the importance of blockchain-enabled energy trading for cyberphysical systems this article provides a thorough overview of Peer-to-Peer (P2P) energy trading, and the utilization of blockchain to enhance efficiency and overall performance of the decentralization, scalability and security of systems. Three blockchain-based energy trading models are proposed to overcome the technical challenges and market barriers, and enhance the adoption of this disruptive technology.

\end{abstract}


\section{Introduction}

The global economy is experiencing a significant growth of businesses with increasing demands for energy. Global energy consumption increased by 2.1\% in 2017, and according to 2019 US Energy Information Administration (EIA) projects the expected growth in Asia by 2050 will be 50\%. Since it is unlikely these demands will be met by conventional energy systems, the world is moving toward renewable energy systems (RES) supported by cyberphysical technologies \cite{muller2017rise}. In addition, climate changes increase the need for reduced dependency on conventional resources, several nations are aiming at increasing the dependency on renewable energies. Owing to this, there has been an increase in the share of renewable energy towards the total energy of the world overall. The shift from conventional energy sources to renewables is clear, as revealed in 2018 when 30\% of the UK gross electricity consumption was generated by RES (According to Green Match report on Renewable Energy in the United Kingdom). This transformation has led to the emergence of prosumers (producers and consumers) which are capable of both generating and consuming energy. Many households and buildings are now equipped with solar panels that can contribute to the overall energy needs. Power systems need to adapt to renewable energy so they can operate efficiently and sustainably, and promote the concept of smart grids to support power systems. 

The emergence of prosumers and smart grids creates opportunities for energy trading transactions that can be done between entities such as prosumers, grids and energy storage. As energy is the most critical component for growing the economy, this paradigm shift in energy trading requires a system that is secure, efficient and fosters energy economics. Moreover, the trading systems should become more decentralized, which will provide the most secure market for additional participants. Blockchain is a promising technology that offers a distributed, robust, secure and private framework for P2P energy trading \cite{salil1}. In a recent report, Navigant Research reported that \textit{Energy Blockchain Applications} are expected to have a compound annual growth rate of 67\% over the next decade. The energy sector can deploy blockchain for quick payments, secure energy transactions and privacy \cite{knirsch2018privacy}.

\begin{figure}[h]
\centering
\hfil
\includegraphics[width=\linewidth]{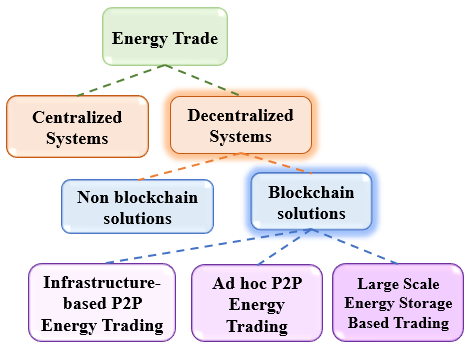}
\caption{P2P energy trading system models}
\label{FLOWCHART}
\end{figure}

Motivated by the need to address these aspects, we first discuss the background of blockchain in P2P energy trading. Then, we enlist open issues in the field and propose three decentralized cyber-physical blockchain-enabled P2P energy trading models, as illustrated in Fig.\ref{FLOWCHART}. We also walk the reader through the common design principle in such environments, analyze the models and conclude the paper with future research directions.

\section{Blockchain and P2P Energy Trading}
\subsection{Blockchain}
The blockchain concept involves development of an immutable and distributed chain-like data structure to securely store data that can be verified as required. Blockchain also offers strong consensus features that allow the ledger to be distributed among all participants. The ledger consists of blocks and each block has a number of transactions; this is known as the block size.

Blockchain uses hash values to keep the chain connected and immutable, and blocks are constructed on top of the hash values. A block is set of approved transactions with a timestamp and hash pointing to the previous block. 
\textbf{Smart Contracts} is a network protocol intended to facilitate, verify or enforce the negotiations or performance of a contract, and smart contracts allow credible transactions without third parties. Energy transactions can use these decentralized smart contracts to develop trust between sellers and buyers. 
\subsection{Blockchain in P2P Energy Trade}
Energy trading is the sale and purchase of specific amounts of energy from the producer to the buyer \cite{Aloqaily2020}. Energy is conventionally traded wholesale, and most of it is transferred from big energy producers (e.g. coal mines, fuel energy generators) to distributors, who then forward the power to commodities such as houses, office buildings and farms
Today, many households and office buildings have solar panels installed on their premises, and this allows them to be producers capable of selling surplus energy to others. This fostered the P2P concept of energy trading, by encouraging consumers to become prosumers (i.e. capable of both producing and selling surplus energy). This means that more energy is available, and consequently reduces overall energy costs.

\begin{figure}[ht]
\centerline{\includegraphics[width=\columnwidth]{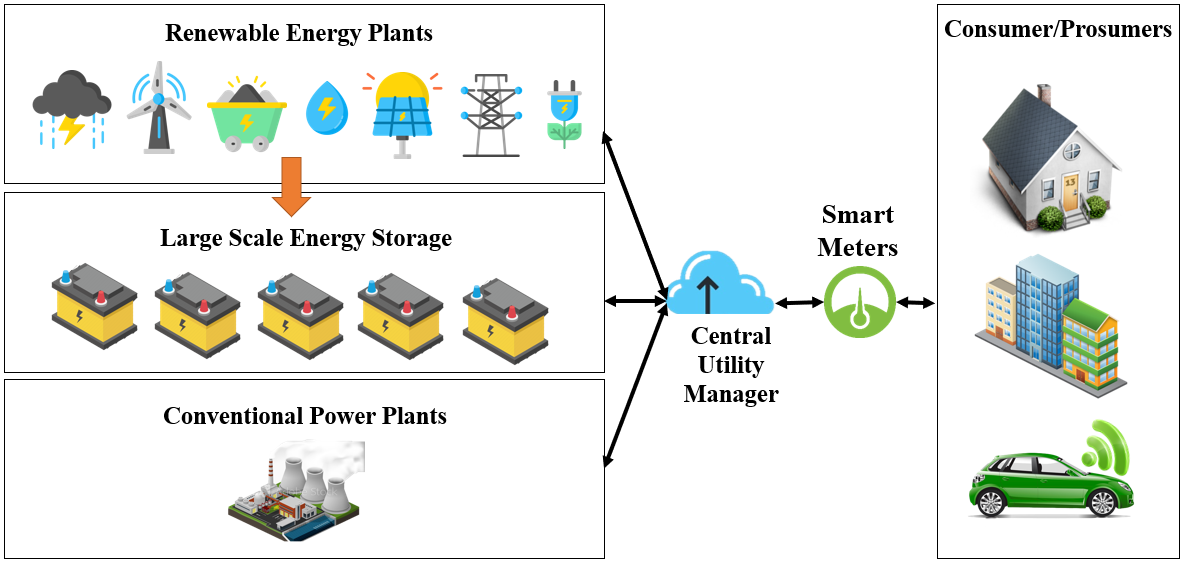}}
\caption{Overview of energy trading}
\label{fig2}
\end{figure}

Energy trading requires management for secure energy trading and supply without any disruptions that can be achieved by a central utility manager, as depicted in Fig.\ref{fig2}. However, a centralized infrastructure can cause issues, including single-point-of-failure, dependency on the central entity and privacy concerns. This supports the transformation from centralized to decentralized models in cyberphysical systems.

Applying blockchain for energy trading has the potential to increase efficiency and security. There are several applications that can contribute to this, including grid management security and accountability, efficient utility billing and P2P energy trading. With regard to security, blockchain can instill trust between parties which removes the dependency on centralized control and, due to the immutability, provides integrity and a high degree of accountability. Moreover, blockchain supports multiple device types and reduces central intermediaries, which also helps reduce energy costs. In April 2016, energy that was generated in a decentralized manner was sold directly between neighbors in New York via a blockchain system. This demonstrated that both energy producers and consumers can execute energy supply contracts without involving third-party intermediaries; thereby effectively increasing speed and reducing transaction costs \cite{pwc2017}. The steps involved in the procedure are as follows:
\begin{itemize}
    \item Electricity is generated.
    \item Consumer buys the energy.
    \item The transaction is updated on the blockchain via a smart meter.
\end{itemize}
Similarly, power ledger trailed blockchain energy trading in rural Australia, allowing commercial buildings to trade excess solar power. (https://www.ledgerinsights.com/power-ledger-blockchain-energy-trading-rural-australia/). A comparable idea is highlighted in \cite{mylrea2017blockchain}.

\subsection{Challenges and Limitations} \label{CNL}

Although blockchain is a promising technology for P2P energy trading, many challenges regarding its adoption have not yet been addressed.

\textbf{Scalability and security}: Blockchain still needs to demonstrate acceptable scalability, efficiency and security for proposed cases. With the growing number of users  of P2P energy trading, implementation of a mature and secure blockchain architecture is critical.

\textbf{Transaction and verification cost:} Transactions become part of the blockchain after initial computation, and this can be a long process. VISA is capable of 20,000 transactions per second, while blockchain can process no more than ten transactions per second. Moreover, there is a transaction charge users must pay to create a transaction, and participating actors require a significant amount of synchronous storage for verification of the blockchain ledger; both these factors increase the overall blockchain cost. As the number of participants in P2P energy trading increases, the number of transactions also grows and can overload the memory requirements for maintaining the ledger, as well as limit the efficiency of blockchain solutions.

\textbf{Development cost:} Another important challenge is blockchain development costs. For example, a transaction verifier needs high computation and network power, which is an extra cost not required for a traditional database system. Blockchain-enabled energy trading also requires costly equipment such as smart meters, which can also increase the cost.

\textbf{Regularization:} The blockchain process is beginning to demonstrate the potential of decentralized energy trading grids. However, the solutions presented in the literature have regulatory issues that include load balancing, integration with central control and coordination with the main grids. For example, if the government wants to manage the power grids it is difficult for them to have greater control and more decentralization. In addition, when using the blockchain the price of energy trading should be regulated.

\textbf{Government Limitations:} The growth of P2P energy trading can lead to governments losing control over energy systems, and they want to retain control to allow for better policies and regularization. In addition, governmental intervention is crucial to supporting the development and commercialization of new technologies, while simultaneously creating markets with advanced energy systems. Blockchain provides P2P and distributed solutions that can limit governmental control, and this is a major challenge to blockchain adoption.


\section{Common Principles}

This section defines principles that are the building blocks of all the proposed models.

\subsection{Design Goals}

Creating blockchain-based infrastructures requires defining a set of technical, regulatory and economic design goals. These \textbf{technical design goals} include::

\begin{enumerate}
    \item Decentralization: Blockchain systems do not depend on centralization, thus they can be used to create energy trading models with no central authority involved.
    
    \item Scalability: The models must be scalable to include the sizes and geographical extents of new participants.
    
    \item Heterogeneity: Due to the different types of devices, such as energy shared between two cars or two houses, devices and systems need to be integrated in the models to support diversity. The system should be transparent and open to anyone who wishes to take part.
    
    \item Intelligence: There are two aspects of intelligence. First, energy should be sold at the optimum (or stable) cost, there should be an intelligent bidding system and the client should have the power of choice. Secondly, energy requires effective management (e.g. demand and response).
    
    \item IoT, Smart Devices and Asset Management: IoT devices are the core components required by blockchain infrastructures. For example, electric vehicles are equipped with IoT devices and sensors that allow communication between autonomous vehicles, allowing them to trade energy between them. 
\end{enumerate}


Inclusion of the blockchain in the energy paradigm means regulatory challenges discussed in the previous section and the presented models have the following \textbf{regulatory design goals}. \textbf{Multi-source Cooperation:} For better collaboration between entities,  privacy and data management systems must be ensured. \textbf{Stakeholder interest:} Stakeholders include governments, national grid operators, microgrid owners, energy suppliers and prosumers. To make optimal use of blockchain-based systems, the interests of all stakeholders must be incorporated in the models. For example, since so many policymakers are shifting to distributed energy, the conventional utilities are facing an uncertain future that can cause delays and disruptions in acquiring distributed energy.

  

The \textbf{economical goals} are as follows:
\begin{enumerate}
    \item \textbf{Pricing:} Since energy trading can decrease the revenues of conventional energy players and disrupt the business of retailers, the pricing of the energy should be standardized.
    
    \item \textbf{Energy markets:} in an economic zone where all parties can sell the energy. One potential pathway is P2P microgrids, which can aggregate prosumer supplies and provide ancillary services such as balancing support. Market competition would benefit from an ancillary service market, and it would  also be important for P2P microgrids as they could realise increased profit for voltage, frequency, restoration, peak load and balancing support.
    
\end{enumerate}
\ 


\subsection{Entities}

As shown in Fig.\ref{fig2}, the entities involved in energy trading are divided into three groups, all of which can have access to smart meters and blockchain.

\begin{itemize}
    \item \textbf{Central utility manager} are government, energy companies and grid owners with the physical and technical infrastructure needed for energy sharing and transfers. They are responsible for regularization and global control of the system.
    
    \item \textbf{Energy generators} are conventional and renewable energy generators with large reservoirs of energy that provide energy for the network for national grid operators, microgrid owners and solar cells and turbine holders.

    \item \textbf{Consumers and Prosumers} are the energy users. Prosumers have extra capabilities that allow them to generate and sell surplus energy to other consumers in the system. Consumers/Prosumers These can be residential houses, electric cars or large buildings.
\end{itemize}



The transaction workflow is divided into three main categories:
\begin{enumerate}
    \item \textbf{Energy transaction} includes all the communication and bargaining that takes place between buyers and sellers.
    
    \item \textbf{Pre-trade communication} includes publishing users’ supply and demands over the network. Different methods can be applied to ensure data privacy and anonymity \cite{wang2019energy}.
    
    \item With \textbf{buyer-seller matching}, the seller makes a bid and matching is conducted to identify the best price for the user. After matching, payment is performed using the transnational settlement method.

\end{enumerate}

\subsection{Payments, Rewards and Demand Response}

To increase benefits for prosumers, consumers can pay with cryptocurrency \cite{kang2017enabling}. This motivates prosumers, since they will receive payments more quickly and have access to investment opportunities. Furthermore, there can be potential reward incentives for more active users. For example, if the prosumer is selling to the government the user can be rewarded in energy points, cryptocurrency or by an adjustment to their bill. 


Real-time control and management play an important role in P2P energy trading. Demand response is the transfer of energy from low demand users to high demand consumers. For example, in the morning, the household requires a lesser amount of energy than the office building. Similarly, at night the situation is reversed. Hence, the energy can be distributed according to the demand.
The limitations of building such a management system and providing smart contract-based demand response system are significant. This control unit is a major aspect of each of the three models defined herein.

\begin{figure}
\centering
\begin{subfigure}{\columnwidth}
    \includegraphics[width=\columnwidth]{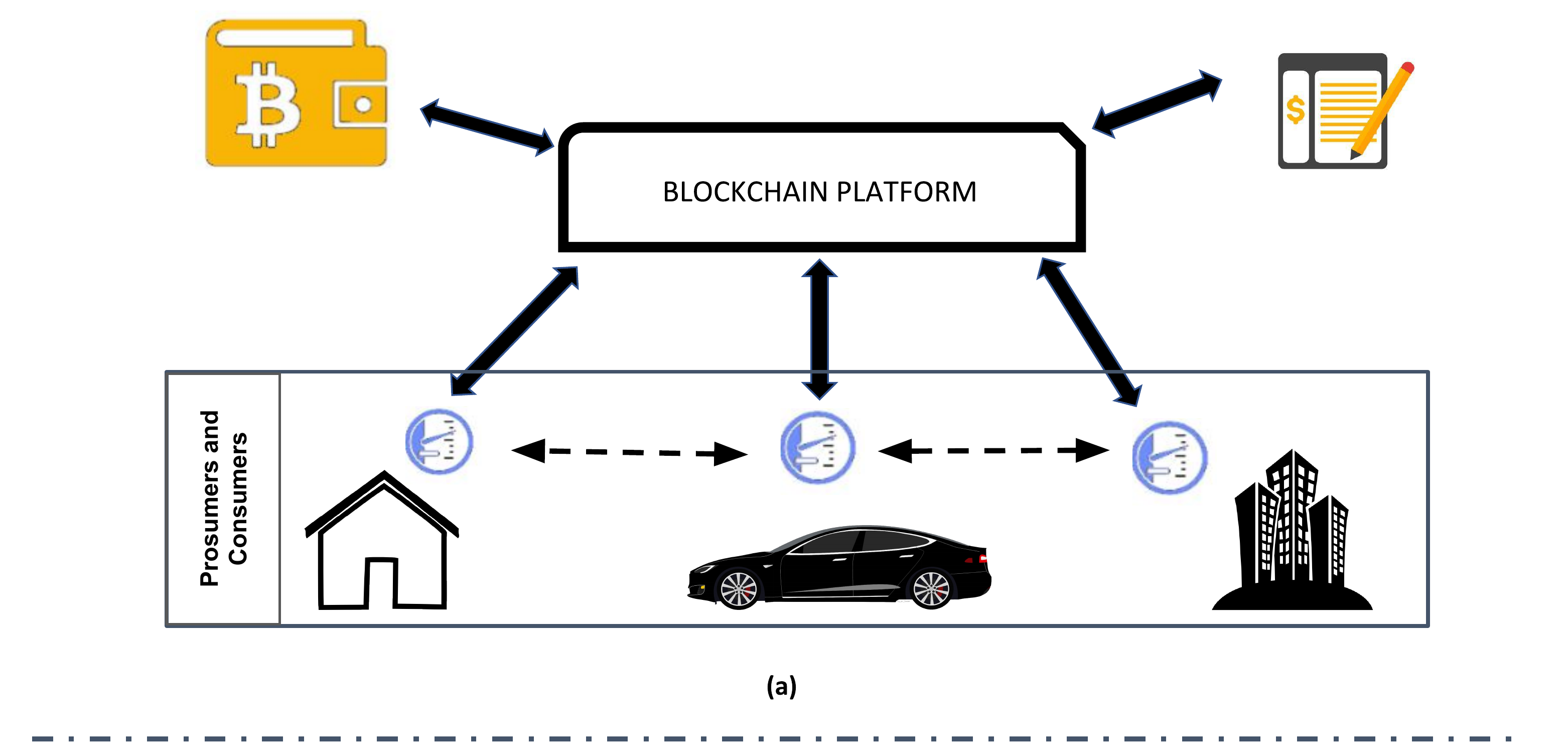}
    \label{CPT}

\end{subfigure}
\begin{subfigure}{\columnwidth}
    \includegraphics[width=\columnwidth]{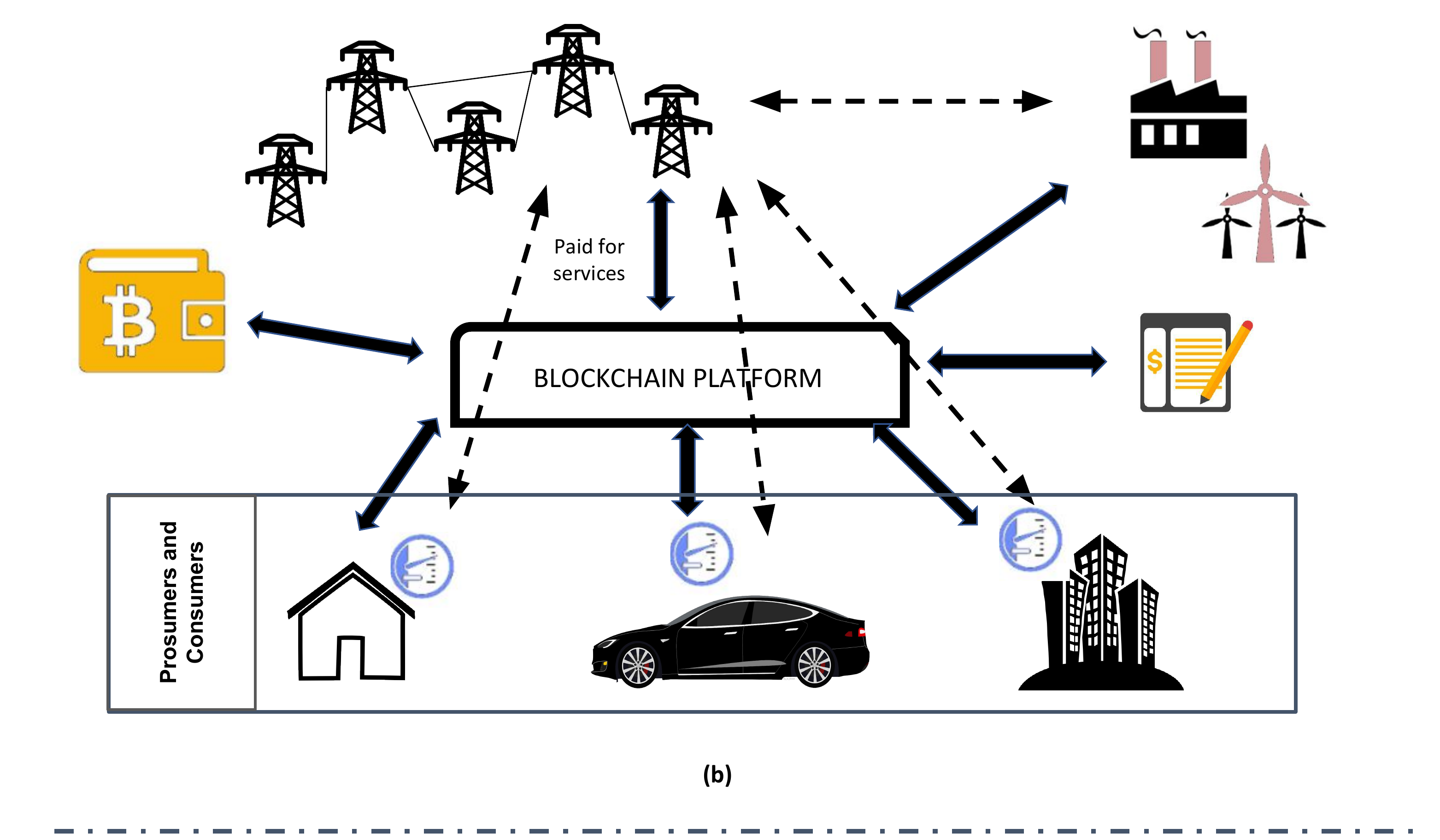}
    \label{CPTA}
\end{subfigure}
\begin{subfigure}{\columnwidth}
    \includegraphics[width=\columnwidth]{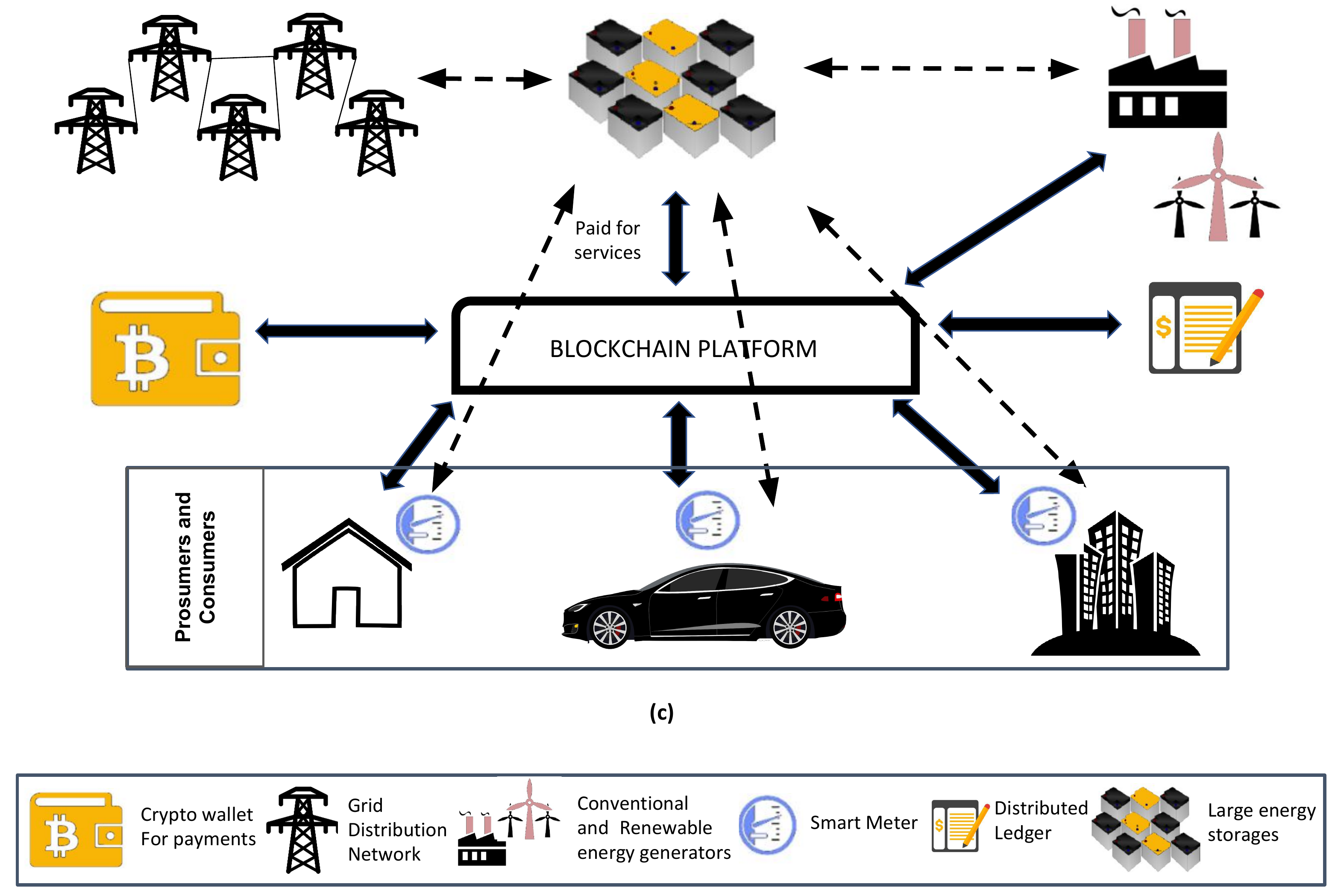}
    \label{CPTB}
\end{subfigure}
\caption{Energy Trade Models: (a) Infrastructure-based P2P Energy Trading, (b) Ad hoc P2P Energy Trading, (c) Large Scale Energy Storage Based Trading}
\label{BGPA}
\end{figure}

\section{Infrastructure-based Energy Trading}
\label{v1}

Traditional energy trading is managed by a centralized organization, but prosumers may not need centralized authority for P2P transactions. Where there are physical means of transferring energy, prosumers can directly communicate with one another. For example, two neighboring houses can connect via cable wire and the energy can be transferred directly. Another example is that two electric cars at the same station can transfer the energy without any intermediary party. 
In the infrastructure-based P2P energy trading model, it is assumed that prosumers are equipped with smart meters and have IoT devices installed to connect to whatever requires the energy (e.g. house, car). As illustrated in Fig 3a, the devices communicate through the blockchain to ensure successful transactions between entities. Another assumption is that prosumers have the physical means to transfer energy between them.

An architecture for pure P2P trading proposed in \cite{salil1} does not need third parties for the negotiations, since the prosumers and consumers communicate directly to create transactions. The architecture here fits this model, as it does not require grid stations during transactions, and if the prosumers have the physical means to transfer energy they can perform the transactions without involving intermediaries. This approach reduces communication from the blockchain layer since the negotiations are done on a separate network. For better security, two transactions have to created within a time-period to be accepted. The solution provides better privacy and security at lower cost and time consumption. 

Brooklyn microGrid \cite{mengelkamp2018designing} in New York is another example that fits our infrastructure-based P2P energy trading model, as it is based on a small number of connected prosumers and consumers (e.g. 5 consumers / prosumers). The prosumers can sell their surplus energy to neighbouring consumers using smart meters and e-wallets, and every member has access to all transactions via self-executing contracts. Users can specify the maximum payment they can afford, and they can prioritize the type of energy required (i.e. conventional or renewable). 

The disconnected micro-grid explained here is premature and has technical challenges, such as adequate physical infrastructure for the energy transfer and the required global reach (e.g. when a seller is not near the buyer but still wants to sell their energy). This can lead to dysconnectivity and isolation of the prosumers, which in turn can cause irregular pricing in different locations. However, the model is efficient for local markets where the traffic in the system is low, and global reach is not required.

\section{Ad hoc P2P Energy Trading}
\label{v2}
Considering the problems with conventional distributed energy trading and the blockchain based model presented here (i.e. infrastructure-based P2P energy trading) energy trading must involve the grids and the infrastructure. Ad hoc P2P energy trading models have local micro-grids integrated with large scale energy producers through blockchain - based platforms (Fig 3b.). This allows consumers to buy energy from another prosumer or from conventional power plants. Due to the immutability and distributed nature of blockchain, the models ensure that all transactions are available to all the prosumers as well as the big energy companies, particularly governments. The platform can be provided by the government, which essentially gives them control over the energy sharing economy. This would mean all involved parties get more value and increased motivation. Since the distributors would be rewarded for these services, Ad Hoc model also provides business opportunities for conventional distributors. 

\begin{table*}
\centering
\caption{Comparison of the energy trade models, $\uparrow$ \text{means comparatively high}, $\downarrow$ \text{means comparatively low}}
\resizebox{\textwidth}{!}{
\begin{tabular}{|l|c|c|c|c|c|}
\hline
        & \multicolumn{3}{c|}{\textbf{Blockchain Solutions}}                                                                                                                                                                                                               & \multicolumn{2}{c|}{\textbf{Non-Blockchian Solutions}}                                                                                                            \\ \hline
\backslashbox{\textbf{Features}} {\textbf{  
Models}}                                           & \textbf{\begin{tabular}[c]{@{}c@{}}Infrastructure-based \\ P2P Energy  Trading\end{tabular}} & \textbf{\begin{tabular}[c]{@{}c@{}}Ad  hoc  P2P\\ Energy  Trading\end{tabular}} & \textbf{\begin{tabular}[c]{@{}c@{}}Large Scale Energy \\ Storage Based Trading\end{tabular}} & \textbf{Centralized}                                                   & \textbf{Decentralized} \\ \hline
\textbf{Entities}                                                                  & Prosumers                                                                                    & \begin{tabular}[c]{@{}c@{}}Prosumers and \\ grid stations\end{tabular}          & \begin{tabular}[c]{@{}c@{}}Prosumers, grids and\\ energy storages\end{tabular}  & \begin{tabular}[c]{@{}c@{}}Prosumers, grids \\and energy storages\end{tabular} & \begin{tabular}[c]{@{}c@{}}Prosumers, grids \\ and energy storages\end{tabular}   \\ \hline
\textbf{Single Point of Failure}        & \multicolumn{3}{c|}{No}                                                                                                                                                                                                                                          & \multicolumn{2}{c|}{Yes}                                                                                                                                          \\ \hline
\textbf{Energy Profile Anonymity}       & \multicolumn{3}{c|}{Yes}                                                                                                                                                                                                                                         & \multicolumn{2}{c|}{No}                                                                                                                                           \\ \hline
\textbf{Decentralization}                                                          & $\uparrow$                                                                                      & $\uparrow$                                                                         & $\uparrow$                                                                         & $\downarrow$                                                                      & $\downarrow$                                                                        \\ \hline
\textbf{Blockchain}                                                                & Smart Contracts                                                                              & Smart Contracts                                                                 & Smart Contracts hyperledger         & -                                                                              & -                                                                                \\ \hline
\textbf{Energy Agreement Verification}  & \multicolumn{3}{c|}{Through blockchain consensus between all nodes}                                                                                                                                                                                              & By central authority                                                           & \begin{tabular}[c]{@{}c@{}}By distributed \\ consensus\end{tabular}              \\ \hline
\textbf{Energy trade}                                                            & Pure P2P                                                                                     & Hybrid P2P                                                                     & Hybrid P2P                                                                     & -                                                                              & Hybrid P2P                                                                      \\ \hline
\textbf{Rewards and Payment}            & \begin{tabular}[c]{@{}c@{}}Cryptocurrency or \\ Energy coin\end{tabular}                     & \begin{tabular}[c]{@{}c@{}}Cryptocurrency or \\ Energy coin\end{tabular}        & \begin{tabular}[c]{@{}c@{}}Cryptocurrency or \\ Energy coin\end{tabular}        & \begin{tabular}[c]{@{}c@{}}Visa/bank \\ Transactions\end{tabular}              & \begin{tabular}[c]{@{}c@{}}Visa/bank \\ Transactions\end{tabular}                \\ \hline
\textbf{Heterogeneity}                                                             & $\uparrow$                                                                                      & $\uparrow$                                                                         & $\uparrow$                                                                         & $\downarrow$                                                                      & $\downarrow$                                                                        \\ \hline
\textbf{Market Approach}                                                           & Limited to local area                                                                        & Limited to local city                                                           & Global Approach                                                                 & Global Approach                                                                & Global Approach                                                                  \\ \hline
\textbf{Trust between trustless parties} & $\uparrow$                                                                                      & $\uparrow$                                                                         & $\uparrow$                                                                         & $\downarrow$                                                                      & $\downarrow$                                                                        \\ \hline
\textbf{Demand Response}                                                           & $\downarrow$                                                                                    & $\downarrow$                                                                       & $\uparrow$                                                                         & \multicolumn{2}{c|}{By Central Authority ($\uparrow$)}                                                                                                               \\ \hline
\textbf{Central Control}                                                           & Minimal                                                                                      & Partial                                                                         & Complete                                                                        & Complete                                                                       & Partial                                                                          \\ \hline
\end{tabular}
}
\label{TABLE1}
\end{table*}

As shown in 
Fig 3b, the structure of this model assumes smart devices are available for both prosumers and energy producers. The architecture is divided into four layers to enhance the development and analysis of the system:

\begin{enumerate}
    \item \textbf{The energy grid} layers include all big companies and stakeholders that can either create energy or are distributors.
    
    \item \textbf{The communication layer} is where all the pre-bargain and communication is done. Here, the buyer initiates a request to buy energy, and the response can be through one of two protocols: matching or biding. The buyer is first matched with sellers  according to the buyer’s preference (e.g. minimum cost seller); the sellers submit their bids and the buyer commits to one  . Another communication then takes place between the seller and the grids to establish the price of the services provided by the distributor.

    \item \textbf{The blockchain layer} is responsible for all transactions, and employs a smart contract-based system that is triggered whenever there is a transaction between two parties. Two separate ledgers are used, and a transaction between the buyer/seller and distributor also makes the system more robust, scalable and efficient.

    \item \textbf{The consumer/prosumer layer} contains all the consumers and prosumers, and manages connections between members and the other layers through IoT devices.
\end{enumerate}

An architecture that fits into our ad hoc P2P energy trading model and uses proof-of-stake based permission blockchain for efficiency is presented in \cite{siano2019survey}. Surplus energy is sold to other network participants via the energy manager, and a transaction controller makes energy bids and offers to buy or sell energy through the architecture; all transactions are retained by blockchain for accounting purposes. With regard to electric vehicles (EV), other research proposes \cite{knirsch2018privacy} allowing EVs to use blockchain to locate a charging station, and the charging stations could bid to charge EVs. This process would help find the best price and location for both EV users and charging stations, while providing privacy and security for the EVs. In this approach, there is no physical infrastructure but still EVs can make benefit from the energy trading.
\section{Large Scale Energy Storage Based Trading}
\label{v3}

Prosumers can take advantage of the architecture available for energy transfers provided by the large scale energy storage entities by selling the energy to the local grids directly (Fig 3c). This would require an agreement between the two parties that is recorded over the blockchain. In this case, the parties would be both prosumers and grids, so this model is under the paradigm of P2P energy trading. 

A model that fits this category \cite{wang2019energyc} provides a functional framework and model of energy distribution for crowdsourced energy systems, and also considers various types of energy trading transactions. The framework enables P2P energy trading at the distribution level, where distribution level asset owners can trade with one another. An operator is responsible for ensuring the transactions do not violate technical constraints. 
The first phase focuses on the day-ahead scheduling of energy generation, and controllable DERs manage the bulk of the grid operations. The second phase is developed to balance hour-ahead or real-time deficit/surplus in energy using monetary incentives. The developed two-phase algorithm supports arbitrary P2P energy trading between prosumers and utilities, which results in a systematic method to manage distribution networks while P2P energy trading is being conducted, while incentivizing crowdsources to contribute to ecosystems. 

This model motivates prosumers to participate in energy trading because there are no infrastructural costs to start selling energy. As well, the prosumer is not responsible for the biding and transactions. Scanergy \cite{scanenergy} research project uses this approach with prosumers assisting to balance supply/demand in return for rewards provided by the grid. 
Prosumer energy exports are recorded by smart meters, and the prosumer is rewarded once its exported energy is used by another consumer. Though this approach can limit the control the prosumer has over the energy trading, it can also decrease management costs since the centralized energy storage systems are responsible for the distribution of the energy to the buyers.

The purpose of these models is to address gaps in the current architecture. Table \ref{TABLE1}, compares the features of blockchain solutions with traditional centralized and distributed architectures, and highlights the blockchain improvements in security, heterogeneity and openness of the systems. There is also less central control and notable efficiency improvements.  The differences between the three blockchain models are summarized according to market approach, payment methods and demand responses.

\section{Open Issues and Future Directions}

The open issues and future directions that we identified in the context of cyberphysical blockchain-enabled P2P energy trading systems are discussed under the following categories:

\textbf{Integration of Energy Distribution without Smart Meters:} Many energy distribution architectures used by prosumers/consumers do not involve smart devices. Since most of the energy trading solutions discussed here assume that prosumers and consumers are equipped with smart devices, a new system can be implemented that can incorporate these architectures with blockchain.s

\textbf{Blockchain as a Black Box:} Most blockchain-enabled solutions use the blockchain as a blackbox. For example, several solutions \cite{salil1}\cite{wang2019energyc} use smart contracts as a blockchain protocol for developing the architecture. This reduces control over the cost and efficiency of the overall architecture, as no changes can be applied to the blockchain used by the smart contracts. Problem-specific blockchains (and consensus) could be implemented for energy trading, rather than using blockchain as a black box. 

\textbf{Blockchain based Prosumer Community Groups (PCG):} Collective trading of energy by a group of prosumers can outperform an individual prosumer with respect to efficiency and reliability of a sustainable energy supply. In addition, an individual prosumer’s energy supply could be too low to compete with traditional energy generators, and could also be unpredictable due to dependencies on climatic conditions. These challenges led to the emergence of PCGs (trade the energy through a community gateway) \cite{rathnayaka2014goal}. As PCGs are centralized in nature, making them robust and secure will require promising technologies like blockchain for development and implementation. 

\textbf{Optimizations:} Demand response and price optimization are still not fully addressed, and further research could include machine learning-based solutions for energy management and cost analysis for P2P energy trading. For example, a real-time billing system that can optimize the energy costs by current and predicted future prices, and charge the user accordingly.  

\textbf{Technology Integration:} A platform is needed to allow the models to work as a single architecture, and change their behaviour according to prosumer requirements. For example, prosumers should be able to sell their energy locally, and directly to large scale energy storage entities.








\section*{Acknowledgement}
This research was partially funded by Koç University - Tüpraş Energy Research Center (KUTEM), and College  of  Engineering,  Al Ain  University, UAE under grant No. ERF-20.

\balance
\bibliographystyle{IEEEtran}
\bibliography{IEEEabrv,main}

\vspace{.1cm} 
\textbf{\textit{Faizan Safdar Ali}} received his BSc. degree in Computer Science from the LUMS, Pakistan in 2017. He is currently an MSc student in Computer Science and Engineering at Koç University, Turkey. \par
\vspace{.1cm}
\textbf{\textit{Moayad Aloqaily (S’12, M’17)}} received the MASc. and Ph.D. degrees in Electrical and Computer Engineering from Concordia University and University of Ottawa, Canada in 2012 and 2016, respectively. He is an assistant professor with Al Ain University, UAE. His current research  interests include  Connected  Vehicles, Blockchain  Solutions, and  Sustainable Energy.\par
\vspace{.1cm}
\textbf{\textit{Omar Alfandi}} received the Ph.D. (Dr.rer.nat.) degree in computer engineering from the Georg-August University of Goettingen, Germany, in 2009. He is an associate professor with Zayed University, UAE.
\vspace{.1cm}

\textbf{\textit{\"Oznur \"Ozkasap (SM'19)}} received the Ph.D. degree in Computer Engineering from Ege University and was a Graduate Research Assistant with the Department of Computer Science, Cornell University, where she completed her Ph.D. dissertation. She is a Professor with Koç University, leading Distributed Systems and Reliable Networks Research Group. She serves as an Area Editor for the Future Generation Computer Systems journal, and is a recipient of The Informatics Association of Turkey 2019 Prof. Aydın Köksal Computer Engineering Science Award.

\end{document}